\documentclass[useAMS,usenatbib]{mn2e}
\usepackage{amssymb} 
\usepackage{verbatim} 

\usepackage{graphicx} 

\renewcommand{\H}{{\cal H}}
\newcommand{\WMAP}{{\slshape WMAP~}}

\newcommand{\cCMBFAST}{{\sc cross\_cmbfast}}

\newcommand{\beam}{B_\ell}	

\newcommand{\LLambda}{{\rm de}}
\newcommand{\mnras}{\mbox{MNRAS}}
\newcommand{\aj}{\mbox{{AJ}}}
\newcommand{\apjl}{\mbox{{ApJL}}}
\newcommand{\prd}{\mbox{{PhRvD}}}

\newcommand{\apj}{\mbox{{ApJ}}}
\newcommand{\prl}{\mbox{{PhRvL}}}
\newcommand{\apjs}{\mbox{{ApJS}}}

\def\gsim{ \,\lower .75ex \hbox{$\sim$} \llap{\raise .27ex \hbox{$>$}} \, }
\def\lsim{ \, \lower .75ex \hbox{$\sim$} \llap{\raise .27ex \hbox{$<$}} \, }

\title[Covariance of DE parameters and sound speed]{Covariance of dark energy parameters 
and sound speed constraints from large HI surveys}
\author[A.~Torres-Rodr\'iguez, C.~M.~Cress \& K.~Moodley]
  {A.~Torres-Rodr\'iguez$^{1}$, C.~M.~Cress$^{2}$, K.~Moodley$^{1}$\\
$^{1}$Astrophysics and Cosmology Research Unit, University of KwaZulu-Natal, Westville, 4000, South Africa\\
$^{2}$Physics Department, University of the Western Cape, Cape Town 7535, South Africa}
\date{\today}

\pagerange{\pageref{firstpage}--\pageref{lastpage}} \pubyear{2006}
	
\def\LaTeX{L\kern-.36em\raise.3ex\hbox{a}\kern-.15em
    T\kern-.1667em\lower.7ex\hbox{E}\kern-.125emX}

\begin{document}

\label{firstpage}

\maketitle

\begin{abstract} An interesting probe of the nature of dark energy is the measure of its
sound speed, $c_s$. We review the significance for constraining sound speed models of dark energy using large neutral hydrogen (HI) surveys with the Square Kilometre Array (SKA). Our analysis considers the effect on the sound speed measurement that arises from the covariance of $c_s$ with the dark energy density, $\Omega_\LLambda$, and a time-varying equation of state, $w(a)=w_0+(1-a)w_a$. We find that the approximate degeneracy between dark energy parameters that arises in power spectrum observations is lifted through redshift tomography of the HI-galaxy angular power spectrum, resulting in sound speed constraints that are not severely degraded. The cross-correlation of the galaxy and the integrated Sachs-Wolfe (ISW) effect spectra contributes approximately $10$ percent of the information that is needed to distinguish variations in the dark energy parameters, and most of the discriminating signal comes from the galaxy auto-correlation spectrum. We also find that the sound speed constraints are weakly sensitive to the HI bias model. These constraints do not improve substantially for a significantly deeper HI survey since most of the clustering sensitivity to sound speed variations arises from $z \lsim 1.5$. A detection of models with sound speeds close to zero, $c_s \lsim 0.01,$ 
is possible for dark energy models with $w\gsim -0.9$.

\end{abstract}


\section{Introduction} \label{sec:Introduction}

The observed acceleration of the cosmic expansion has challen\-ged our understanding of the composition and evolution of the universe. Evidence supporting the idea that about two-thirds of the energy in the universe is in the form of dark energy driving this acceleration has arisen from observations of type Ia supernovae \citep*{Riess98, Perlmutter99} and from a combination of large-scale structure \citep{TegmarkLSS} and cosmic microwave background (CMB) observations \citep*{SpergelWMAP3}.

Theoretical explanations of the observed acceleration can broadly be classified into three 
categories \citep*{BeanCarrol05}. Firstly, there are models that remove the need for
a new exotic component by seeking alternatives to dark energy, for example, by modifying gravity on cosmological scales (e.g.~\citealt*{Dvali00,Deffayet01,Deffayet02,Carroll05,Capozziello06}, also see \citet*{Bean07} and references therein). Alternatively, if one assumes the validity of general relativity and interprets the observational evidence as that for dark energy, the simplest theoretical explanation is that of the cosmological constant ($\Lambda$), whose main difficulty is the dramatic inconsistency between its measured value and the predicted value from quantum field theory \citep{Weinberg89,Carroll01}. The third category includes dynamical dark energy theories such as quintessence (e.g.~\citealt*{PeeblesRatra88,RatraPeebles88,Wetterich88, FerreiraJoyce97, CaldwellDaveSt98}, also see \citet*{Linder07} and references therein) and \mbox{k-essence} models \citep*{Armendariz00,Armendariz01,Chiba00,Chiba02}.

In dynamical dark energy theories, dark energy is typically modelled as a scalar field, with
k-essence differing from quintessence in that it has a non-canonical kinetic term in
the Lagrangian. The mechanism by which quintessence comes to dominate at later times,
generating the accelerated expansion, is not clearly identified, leading to the so-called
coincidence problem. The k-essence models were proposed to explain the late-time
domination in a more natural way. Recently it has been claimed that k-essence models are not
physical because in order to solve the coincidence problem in these models the speed of sound of dark energy, $c_s^2=\delta p/\delta\rho$, has to be greater than unity in some epoch \citep*{Bonvin06,Bonvin07} which violates causality. Other authors, however, (e.g.~\citealt*{Kang07,Babi07}) have argued that superluminal sound speed propagation does not lead to causality violation. There also exist k-essence models in which the sound speed is always less than unity but these do not appear to solve the coincidence problem \citep{Scherrer04,Scherrer06}.

While the viability of k-essence models is debated, it is nevertheless useful to consider
experiments which could distinguish between this type of dark energy model
and quintessence. The sound speed for quintessence is always unity so that, as in the case 
of the cosmological constant, the quintessence field is expected to have no significant 
density fluctuations within the causal horizon, consequently it should contribute little to 
the clustering of matter in large-scale structure \citep{FerreiraJoyce98}.  In 
k-essence models, however, the sound speed is not unity and dark energy can cluster, thereby 
affecting the growth of large-scale structure. The detection of a signature of sound speed 
in the integrated Sachs Wolfe (ISW) effect as well as in the clustering of matter, can 
therefore provide valuable insight into the nature of dark energy.

Many experiments have been proposed to probe the properties of dark energy. While many have pursued constraining its equation of state, $w =p/\rho$, for example, using cluster counts surveys \citep*{Haiman01,Weller02}, baryon acoustic oscillations \citep*{BlakeGlazebrook03,HuHaiman03,Cooray04}, weak lensing \citep*{Huterer02,TakadaJain04,SongKnox04} and type Ia supernova experiments (e.g.~\citealt*{WellerAlbrecht02,Nesseris05}), only a few have considered the prospects for sound speed detection \citep*{HuScran04,Dedeo04,WellerLewis03,Hannestad05,BeanDore04, Corasaniti05}. In \citet[hereafter TRC07]{Paper1}, we studied the potential for
combining data from the Square Kilometer Array (SKA\footnote{\texttt{http://www.skatelescope.org/}}) with forthcoming 
Planck\footnote{\texttt{http://www.rssd.esa.int/index.php?project=PLANCK}} data to measure the sound speed. Assuming the SKA will provide a redshift survey of 21-cm emitting galaxies over most of the sky out to z\,$\sim$\,2, and a specific model for the evolution of HI in galaxies, we
considered a combination of the galaxy auto-correlation and the galaxy-CMB cross-correlation
in several cosmic epochs to trace the evolution of clustering. This evolution is somewhat
sensitive to the sound speed of dark energy in the low-sound-speed regime. The
cross-correlation signals are dominated by contributions from the ISW effect and thus provide
a probe of the evolution of potentials on large scales where clustering of dark energy could
be most noticeable. We found that for models with constant $w$, given high precision 
measurements of other cosmological parameters, it was possible to measure the effects 
of a sound speed that is much smaller than unity.

In this paper we extend our analysis to study the covariance of the sound speed with
other dark energy parameters, particularly $w_0, w_a$ and $\Omega_\LLambda$. Since there are
degeneracies in the parameter space when using the ISW effect and the clustering of HI galaxies as observables, it is important to consider the effect that uncertainties in these parameters have
on the sound speed forecasts. For example, increasing the density of dark energy produces a 
larger suppression of the gravitational potential but this can be compensated by making the 
nature of dark energy more like matter by reducing its sound
speed. Similarly, a lower value of $w$ boosts the ISW effect at recent
times but this can be compensated by reducing the sound speed of dark energy. 
We implement a Fisher matrix analysis to 
study the covariance between the dark energy parameters and calculate the significance
for ruling out dark energy models with constant sound speed much less than that
of canonical models. In particular, many k-essence models have
low sound speeds for most of the period between last scattering and the present epoch 
\citep{Erickson}. 

We also study the sensitivity of the sound speed constraints to different models 
for the HI bias, which has a direct impact on the galaxy power spectrum. Recently there has 
been discussion of radio telescopes other than the SKA which could carry out deep HI-galaxy 
redshift surveys over the same area of sky, and which could potentially be built before the 
multi-purpose SKA. We investigate how constraints on the sound speed improve for a much deeper HI-galaxy redshift survey.

The outline of the paper is as follows. In \S \ref{sec:theory} we present the 
ISW and galaxy observables.  In \S \ref{sec:SKA} we revise the properties of our model
HI redshift distribution and the HI bias. In \S \ref{sec:Method} we describe the
statistical method used to derive parameter constraints 
and its numerical implementation. Finally, the results are presented and discussed in \S
\ref{sec:Results} with conclusions drawn in \S \ref{sec:Conclusions}.

\section{Theory}\label{sec:theory}

\subsection{Observables}\label{sec:Overview}

We first review the ISW and HI-galaxy clustering observables which are used to discriminate 
dark energy models. More details can be found in TRC07.

The fluctuations in the matter distribution of the large-scale structure can 
be expressed in terms of the projected fractional source count of the mass tracer
\begin{eqnarray} \frac {\delta N}{N_0}(\hat{n}) &=& \int_{0}^{z} b_{HI}(z)\,
\frac{d\tilde{N}}{dz}\, \delta_m (z,\hat{n})\, dz \nonumber \\ &=& \delta_m^
0(\hat n) \int_{0}^{z} b_{HI}(z)\,\frac{d\tilde{N}}{dz}D(z)\, dz\,, \label{galfield}
\end{eqnarray} 
where \emph{b$_{HI}$} is the linear bias parameter of the HI-galaxy population, $\delta_m$ is
the matter overdensity ($\delta_m^0\equiv\delta_m(z=0)$), $d\tilde{N} / dz$ is the normalised
redshift distribution of HI galaxies and $D(z)$ is the linear growth of matter
fluctuations given by $D(z) = \delta_m(z) /\, \delta_m^0$.

The Fourier modes of the temperature fluctuations originating from the ISW effect
are expressed as the change in the gravitational potential over conformal time (or comoving
distance, $r$) integrated from today to the epoch of de-coupling
\begin{eqnarray} 
\frac {\delta T}{T_0}(k) &=& -2 \int_{0}^{r_{dec}} dr\, \Phi'(r,k) \label{iswfield} \nonumber \\
&=& \frac{3H_0^2\Omega_m}{c^2 k^2}\, \delta_m^0(k) \int_0^z \frac{dg(z)}{dz}\,dz\,,
\end{eqnarray}
where $H_0$ and $\Omega_m$ are, respectively, the value of the Hubble
constant and the matter density parameter today, $\Phi$ is the Newtonian
gravitational potential, $\delta_m^0(k)$ is the Fourier transform of the
matter distribution and the prime denotes derivatives with respect to
comoving distance. The dominant contribution to the ISW effect comes
from the CDM perturbations. This allows us to express the evolution of
the gravitational potential as the change of the linear growth
suppression factor, $g(z)\!=\!(1+z)D(z),$ via the Poisson equation.

We can express the galaxy auto-correlation, CMB auto-correlation and 
galaxy-CMB cross-correlation in harmonic space via their respective angular power spectra as
\begin{eqnarray}
C_\ell^{gg} &=& 4\pi \int \frac {dk}{k}\; \Big \langle \big|\frac{\delta
N}{N_0}(k)\big|^2\Big\rangle\, j_\ell^2(k r) \nonumber \\
&=& 4\pi  \int_{0}^{\infty} \frac {dk}{k}\, [f_\ell^N(k)]^2 \Delta_m^2(k)\, , \label{autogal} \\
C_\ell^{TT} &=& 4\pi \int \frac {dk}{k}\; \Big \langle \big|\frac{\delta
T}{T_0}(k)\big|^2 \Big \rangle \,j_\ell^2(k r) \nonumber \\
&=& 4\pi  \int_{0}^{\infty} \frac {dk}{k}\, [f_\ell^T(k)]^2 \Delta_m^2(k)\, , \label{autotemp} \\
C_\ell^{gT} &=& 4\pi \int \frac {dk}{k}\; \Big \langle \frac{\delta
N}{N_0}(k) \frac{\delta T}{T_0}(k) \Big \rangle \,j_\ell^2(k r) \nonumber \\
&=& 4\pi  \int_{0}^{\infty} \frac {dk}{k} f_\ell^N(k)f_\ell^T(k) \Delta_m^2(k)\, , \label{cross}
\end{eqnarray}
where $j_\ell(k r)$ is the spherical Bessel function and $\Delta_m^2(k)=k^3 P_\delta(k)/2\pi^2$ is the logarithmic matter power spectrum today with $P_\delta(k)=\langle|\delta_m^0(k)|^2\rangle$. The functions $f_\ell^N(k)$ and $f_\ell^T(k)$
correspond to the weight functions for the HI survey and the ISW spectra respectively. 
They depend on both the redshift distribution of the galaxy
selection function and the rate of change of the gravitational potential according to the
cosmological model of dark energy. From Eqs.~(\ref{galfield}, \ref{iswfield}) the weight
functions are defined as 
\begin{eqnarray}
f_\ell^N(k) &=& \int_0^z b_{HI}(z)\frac{d\tilde{N}}{dz}D(z)\,j_\ell(k r(z))\,dz\,,
\label{filtergal} \\
f_\ell^T(k) &=& \frac{3H_0^2\Omega_m}{c^2 k^2}\int_0^z \frac{dg(z)}{dz}\,j_\ell(k
r(z))\,dz.   \label{filterisw}
\end{eqnarray}

\subsection{The effect of dark energy on the observables}\label{sec:effect}

The background expansion of the Universe is affected by the density of the matter component
and by the equation of state parameter, $w$, through the Hubble parameter, $H(z)$. This modifies the comoving distance, $r$, and has a direct effect on Eqs.~(\ref{autogal}--\ref{cross}).

In a general fluid description for the density perturbations of dark energy
(e.g.~\citealt{BeanDore04}), the evolution of the fluctuations is characterised by both the
equation of state and the speed of sound. In the frame where cold dark matter (CDM) is at rest, 
the evolution of the density and velocity perturbation of a general matter component 
(denoted by subscript
\textit{i}) is given by \citep{WellerLewis03} 
\begin{eqnarray}
\delta_i' + 3\H(\hat{c}_{s,i}^2-w_i)(\delta_i +3\H(1+w_i)v_i/k)+ \nonumber \\
(1+w_i)kv_i + 3\H w'v/k = -3(1+w_i)h' \label{evoldens}  \\
v_i' + \H(1-3\hat{c}_{s,i}^2)v_i = k \hat{c}_{s,i}^2 \delta_i/(1+w_i) \, ,\label{evolvel}
\end{eqnarray}
where $\H$ is the conformal Hubble parameter, $v_i$ is the velocity, $h'$ the synchronous
metric perturbation \citep{EvoEqu} and the circumflex \mbox{$(\,\hat{}\,)$} means the sound
speed is defined in the rest frame of the dark energy component. Eq.~(\ref{evoldens}) includes the variations of $w$ with respect to conformal time. We consider models of dark energy with a slowly varying equation of state
parameterized \citep{Linder03} as a function of the scale factor, $a$: 
\begin{equation}
w(a)=w_0+(1-a)w_a . \end{equation}

We note that $\Omega_\LLambda$ only appears directly in the background expansion, unlike
$w(a)$ which, together with $c_s$, has a direct contribution to the density evolution. As we
will see later, this has an effect on the covariance of these parameters.

\section{The HI-galaxy survey}\label{sec:SKA}

In TRC07 we presented the motivation for using an experiment like the SKA to complete
a large HI survey. There are some constraints on the distribution of HI gas at high redshift from Ly-$\alpha$ absorption studies \citep{Peroux03} but there remains a great deal of uncertainty in the HI selection function at high redshift. As our reference model, we adopt the model `C' in \citet{AbdallaRawlings05} and investigate the impact of changing the selection function by, firstly, changing the integration time and, secondly, changing our assumption about the evolution of the HI bias.

\begin{figure}
\includegraphics[width=8.5cm,keepaspectratio]{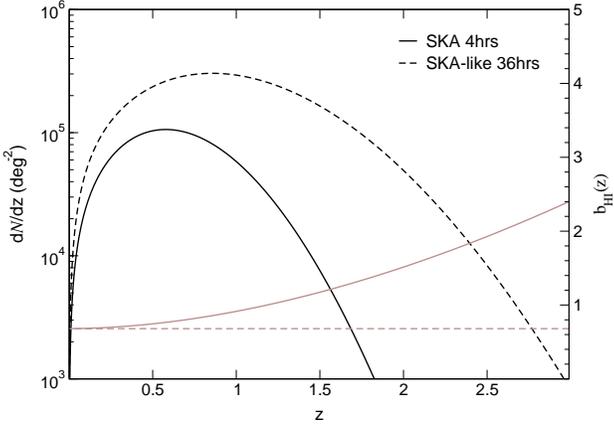}
\caption{SKA redshift distributions of HI galaxies and models for the HI bias parameter used.
An integration time for a single pointing of 36 hours (black dashed line) would reach
a depth of $z=3$. Taken from \citet{Basilakos07}, the two models for the HI bias (in grey) only
differ greatly at high redshift. A constant bias model (grey dashed line) 
is unlikely to be realistic.} \label{dndz_bias}
\end{figure}

For a single-pointing survey, a field of view (FOV) frequency dependence of $\nu^{-2}$ and a detection threshold of $S/N=10$, the number of HI galaxies per square degree, per redshift interval and at redshift $z$ is expected to be reasonably approximated by
\begin{equation}
\frac{dN}{dz}={\cal A}\, z \exp\Big(\frac{-(z-z_c)^2}{2\,\sigma_z^2}\Big)\,,
\end{equation}
where the fitting parameters ${\cal A}$, $z_c$ and $\sigma_z$ depend on the integration time.
We would like to investigate the performance of two different surveys: one of 4 hours of
integration time per pointing (as in TRC07) and an ultra deep survey, using 36 hours of
integration. We assume that both surveys are completed by an SKA-like experiment and 
cover the same area of sky but that the latter survey takes nine times longer to complete. 
Both survey selection functions are depicted in Fig.~\ref{dndz_bias}.

The selection function of our mass tracer is then \mbox{$\phi(z) = b_{\rm HI}(z)\times{\rm
d}\tilde N/dz$}, where the tilde represents the normalised distribution over the redshift of
interest. In order to increase the significance of distinguishing models, we divide the
galaxy selection function into several redshift bins of width $\Delta z = 0.2$ up to $z_{\rm
max}=2$ (or $z_{\rm max}=3$ for the deeper survey), the width of redshift bin being chosen to
minimise shot noise from a small number of galaxies in narrow bins and smearing of 
the gravitational potential along the line of sight in wide bins. We consider the 
galaxy and the ISW-galaxy power spectra in each bin as independent measurements.

From Eq.~(\ref{filtergal}), we see that the bias factor has a direct impact on the galaxy-galaxy
spectra. The effect of the bias model on the sound speed constraints is not straightforward 
to see, so we investigated this further by using different bias models. 
We have used two models for $b_{HI}$ based on the studies of
\citet{Basilakos07} (and references therein). In the first model we make the simple but
unrealistic assumption that the HI bias does not evolve with redshift and that it remains
constant at the value measured locally by HIPASS \citep{Barnes01} in the concordance model, i.e., $b_{HI}=0.68$ \citep{Basilakos07}. More realistically, a second HI bias model considers the transformation of gas into stars and the
gas consumption as a function of redshift. In this model HI galaxies are more biased at high 
redshift relative to the local value. Under the linear perturbation theory,
\citet{Basilakos07} derive the HI bias evolution to high redshift for a $\Lambda$CDM model,
which is shown in  Fig.~\ref{dndz_bias} together with the constant bias model. 
We note that an independent measure of the bias of
HI-selected galaxies as a function of redshift will be possible using redshift-space
distortions in the survey.

\section{Statistical Method} \label{sec:Method}

\subsection{Numerical implementation}

We compute the relevant power spectra in Eqs.~(\ref{autogal}--\ref{cross}) using a modified
version of the \cCMBFAST\footnote{{\ttfamily http://www.astro.columbia.edu/$\sim$pierste/ISWcode.html}} code. These modifications include the addition of our HI-galaxy redshift distribution (see \S \ref{sec:SKA}) as well as the bias factor. The resulting selection function is split into different redshift bins for which the relevant power spectra are calculated.

We choose fiducial models that give an angular diameter distance to recombination fixed to the value from \WMAP observations \citep{SpergelWMAP3}
and let the dark energy parameters $\Omega_{\LLambda}$, $w_0$, $w_a$ and $c_s^2$ vary around
the fiducial model. Unless otherwise indicated, the dark energy fiducial model is: $\Omega_\LLambda=0.705$, $w_0=-0.8$, $w_a=0$ and $c_s=1$. In addition to the dark energy parameters, we assume a physical matter density of $\Omega_mh^2=0.126$, a physical \mbox{baryonic} density of $\Omega_bh^2=0.0223$, an optical depth to reionization of $\tau=0.09$, a primordial power spectrum amplitude of $\Delta_{\mathcal R}^2=2.02\times10^{-9}$ and a scalar spectral index of $n_s=0.951$ \citep{SpergelWMAP3}.

\subsection{Forecasts for sound speed detection} \label{sec:fm}

In order to measure the significance of a detection of a sound speed model we utilise 
the Fisher information matrix. The Fisher matrix is defined as
\begin{equation} 
F_{\alpha \beta} = -
\left\langle \frac{ \partial^2 \ln \mathcal{L} }{\partial \theta_\alpha
\partial \theta_\beta} \right\rangle,
\end{equation} 
where $\mathcal{L},$ the likelihood, is the probability of observing the data set 
$\{ x_1, x_2, ...\}$ for a given cosmological parameter set $\{\theta_1, \theta_2, ...\}$. 

The Fisher matrix method allows us to forecast how well a survey will perform in 
constraining a
set of cosmological parameters, by providing the minimum systematic uncertainty in each model
parameter that is to be fit by the future survey data. Under the assumption that the individual 
parameter likelihoods approximate a Gaussian distribution, the information contained in the 
angular power spectra can be written as \citep*{Heavens}
\begin{equation}
F_{\alpha \beta} =  f_{\rm sky} \sum_\ell { (2\ell+1) \over 2}
\,{\rm Tr}[ 
{\bf D}_{\ell\alpha}  \tilde{\bf C}_{\ell}^{-1}
{\bf D}_{\ell\beta}  \tilde{\bf C}_{\ell}^{-1}
]\,,
\label{eqn:Fisher}
\end{equation}
where the sum extends over multipoles $\ell$, 
$f_{sky}$ is the amount of sky covered by the survey, 
$\tilde{\bf C}_{\ell}$ is the data covariance
matrix, and ${\bf D}_{\ell\alpha}$ is the matrix 
of derivatives of the angular power spectra 
\begin{equation} 
{\bf D}_{\ell\alpha} = {\partial {\bf C}_\ell \over \partial \theta_\alpha}\big|_{\theta_\alpha={\rm fid}}\,
\label{eqn:derivs}
\end{equation}
with respect to the parameters, $\theta_\alpha,$
evaluated at the fiducial model.
The data covariance matrix elements include the angular power spectra plus the noise terms.
For the CMB, the noise contribution to the temperature measurement
depends on the beam window function and the pixel noise of the experiment
\begin{equation} \label{eq:cmbnoise}
\tilde C^{TT}_{\ell} = C^{TT}_{\ell} + w_T^{-1}\beam^{-2}\,,
\end{equation}
where $\beam$ is the window function of the Gaussian beam and
$w_T^{-1} = \sigma_{\rm p}^2~\theta_{\rm beam}^2$ is the inverse noise weight with 
$\theta_{\rm beam}$ and $\sigma_{\rm p}$, respectively, the
beamwidth and noise per pixel in a given frequency band. 
We consider a Planck-like CMB experiment that measures temperature anisotropies in
two high frequency bands, $143$ and $217$ GHz. The details of these parameters for Planck can
be found in e.g.~\citet{Rocha} and are not presented here. As shown in TRC07, the contribution
from the CMB spectrum 
to the detection significance is much less important compared to the
cross-correlation and galaxy auto-correlation spectra. For the galaxy field, the source of noise
comes from Poisson fluctuations in the number density
\begin{equation}\label{eq:Cgg}
\tilde C_\ell^{gg}= C_\ell^{gg} + 1 / \tilde n^z_A \,,
\end{equation}
where $\tilde n^z_A$ is the galaxy number per steradian in the redshift bin of interest.

We construct our data covariance matrix by combining the observables as follows
\vspace{0.3cm}
\begin{equation}
	\tilde{\mathbf C}_{\ell} \equiv \left( \begin{array}{ccccc}
		\tilde C_\ell^{TT}  & C_\ell^{gT, 1} & C_\ell^{gT, 2} &\dots & C_\ell^{gT,10} \\
	\noalign{\medskip}
		C_\ell^{gT,1} & \tilde C_\ell^{gg,1} & 0 & 0 & 0\\
	\noalign{\medskip}
		C_\ell^{gT,2} & 0 & \tilde C_\ell^{gg,2} & 0 & 0\\
		\vdots & 0 & 0 & \ddots & 0 \\
	\noalign{\medskip}
		C_\ell^{gT,10} & 0 & 0 &0&\tilde C_\ell^{gg,10}\\
	\end{array}
	\right)   ,
\end{equation} \vspace{0.1cm}
where the superscript number refers to the redshift bin and the 
zeros indicate no
cross-correlation between the galaxy power spectra from different redshift bins. We have 
also assumed that there is no correlated noise between the galaxy and CMB power spectra.

Information about constraints on the cosmological parameters is contained in the derivatives
in Eq.~(\ref{eqn:Fisher}). We calculate these derivatives for each multipole by fitting a
curve through the power spectra values as they vary as a function of the dark energy
parameters $w_0$, $w_a$ and $\Omega_\LLambda$. We then numerically calculate the slope at the
fiducial point. We require a precision that is less than one percent in order to measure the
small variations in the cross-correlation and auto-correlation spectra. For this reason, we
look at changes in the parameters that are large enough to allow a polynomial curve fitting.

\begin{figure}
\includegraphics[width=9cm,keepaspectratio]{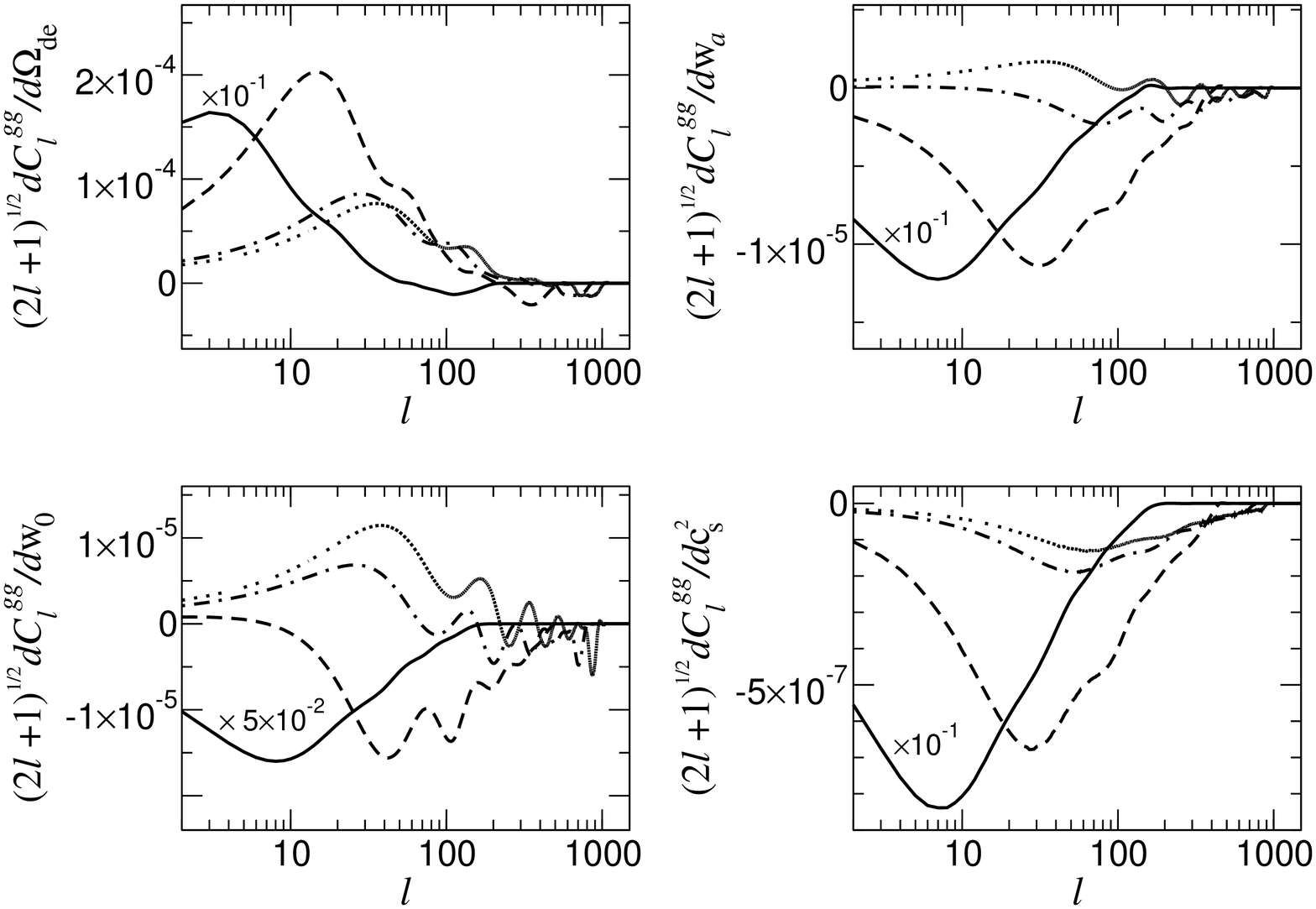}
\caption{Derivatives of the galaxy auto-correlation power spectra with respect to the dark
energy parameters $\Omega_\LLambda$, $w_a$, $w_0$ and $c^2_s$ in redshift bins $0.0<z<0.2$ (solid),
$0.4<z<0.6$ (dashed), $1.0<z<1.2$ (dot-dashed) and $1.4<z<1.6$ (dotted). The fiducial model used is
$w_0=-0.8$, $w_a=0$, $\Omega_\LLambda=0.705$ and $c_s^2=1$.} 
\label{fig:dCgg}
\end{figure}

\begin{figure}
\includegraphics[width=9cm,keepaspectratio]{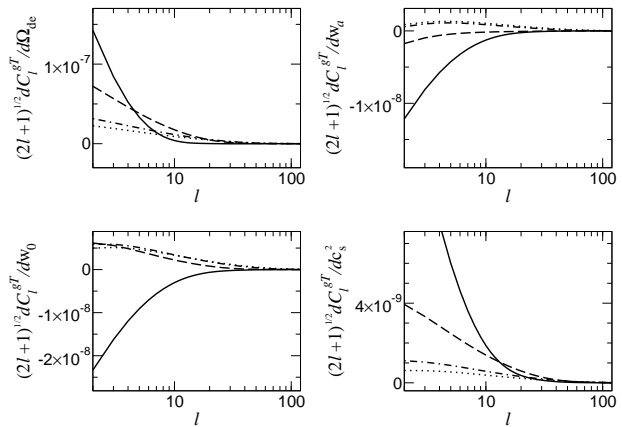}
\caption{Derivatives of the galaxy-ISW cross-correlation power spectra with respect to the 
dark energy parameters $\Omega_\LLambda$, $w_a$, $w_0$ and $c^2_s$ in redshift bins $0.0<z<0.2$ (solid), $0.4<z<0.6$ (dashed), $1.0<z<1.2$ (dot-dashed) and $1.4<z<1.6$ (dotted). The fiducial model is as in Fig.~\ref{fig:dCgg}.} \label{fig:dCgT}
\end{figure}

We present the derivatives of the auto-correlation and cross-correlation power spectra with 
respect to the dark energy parameters in Fig.~\ref{fig:dCgg} and Fig.~\ref{fig:dCgT}, respectively, 
for a series of redshift bins. We have multiplied the derivatives by the factor $(2\ell+1)^{1/2}$ 
to make the contribution of this product to the Fisher information matrix as a function of 
$\ell$ more evident. We note that the galaxy auto-correlation spectra contribute most of the 
signal in distinguishing sound speed models. This is due to the larger amplitude of the 
galaxy auto-correlation derivatives with respect to the dark energy parameters as compared to the 
cross-correlation derivatives. As we will see in the next section the cross-correlation spectra only 
add about ten percent of the discriminating signal. The cross-correlation signal originates from the 
large-scale ISW effect, thus the information from the cross-correlation spectrum is mainly available 
at low multipoles, $\ell \lsim 20$, as can be seen in Fig.~\ref{fig:dCgT}.

The power spectrum derivatives also provide insight into the covariance between the sound speed and
other dark energy parameters. We observe in Fig.~\ref{fig:dCgg} that the derivative with respect to 
$\Omega_\LLambda$ is two orders of magnitude larger than the derivative with respect to the sound 
speed. The larger amplitude and distinct shapes of the $\Omega_\LLambda$ derivatives compared to the 
sound speed derivatives in different redshift bins breaks the degeneracy between 
$\Omega_\LLambda$ and $c_s$. The amplitude of the change in the galaxy auto-correlation and 
cross-correlation spectra due to a varying equation of state ($w_0$ and $w_a$) is smaller than 
the change due to the dark energy density but still an order of 
magnitude larger than that caused by variations in $c_s$. In addition the characteristic shapes 
of the auto-correlation and cross-correlation derivatives with respect to $w_0$ and $w_a$ 
in different redshift bins allow the degeneracy between $w_0, w_a$ and $c_s$ to be lifted.

We extend the sum in Eq.~(\ref{eqn:Fisher}) up to $\ell_{\rm max}=1500$ to
include the galaxy auto-correlation signal from the highest redshift bins, which is projected
onto small angular scales.

\subsection{A quantitative approach}

We have seen how the use of a survey split in redshift bins helps lift the
degeneracies that exist between dark energy parameters. We wish to
quantify the effect on the sound speed measurement by marginalising over $w_a$, $w_0$ and
$\Omega_\LLambda$. The full Fisher matrix for this experiment includes matrix elements for all
parameters
\vspace{0.1cm}

\begin{equation}
	\bf F = \left( \begin{array}{cc}
		F_{cc}  & \bf B  \\
	\noalign{\medskip}
		{\bf B}^{T} & \bf A \\
	\end{array}
	\right),
\end{equation} 
where
\vspace{0.1cm}
\begin{equation}
	\bf A = \left( \begin{array}{cccc}
		F_{w_aw_a}  & F_{w_a w_0} & F_{w_a \Omega_\LLambda} \\
	\noalign{\medskip}
		F_{w_a w_0} & F_{w_0 w_0} & F_{w_0 \Omega_\LLambda}\\
	\noalign{\medskip}
		F_{w_a \Omega_\LLambda} & F_{w_0 \Omega_\LLambda}	& F_{\Omega_\LLambda \Omega_\LLambda}	
	\end{array}
	\right)
\end{equation}
and
\begin{equation}
	\bf B = \left( \begin{array}{ccc}
		F_{w_ac}  & F_{w_0c} &F_{\Omega_\LLambda c}  \\
	\end{array}
	\right).
\end{equation}

We obtain the marginalised Fisher matrix by taking the inverse of $\bf F$ and extracting the
sub-matrix corresponding to the parameters we want to measure. For the sound speed this
is

\begin{equation}
({\bf F}^{-1})_{11} = (F_{cc}-{\bf BA}^{-1}{\bf B}^{T})^{-1},
\end{equation}
and the marginalised Fisher matrix is the inverse of this

\begin{equation}
F_{\rm marg} = F_{cc}-{\bf BA}^{-1}{\bf B}^{T}\,,
\end{equation}
which is a scalar, since we are interested in measuring the sound speed parameter.

In order to compute the derivatives for the Fisher matrix elements involving the sound speed
in Eq.~(\ref{eqn:Fisher}), we note that the effect on the angular power spectra due to changes in the sound speed is only significant for order-of-magnitude variations of $c_s$. Following \citet{HuScran04}, we therefore define the derivative with respect to $c_s^2$ in Eq.~(\ref{eqn:derivs}) as a finite difference

\begin{equation} \label{eqn:cderivs}
{\bf D}_{\ell \,c_s^2} \rightarrow {\bf C}_\ell(c_s^2=1) - {\bf C}_\ell(c_s^2\neq 1)\,,
\end{equation}
noting that, with this definition, the value of $F_{\rm marg}$ can be interpreted as the
significance, $(S/N)^2,$ of a detection of a sound speed model with $c_s^2\neq 1$ 
relative to a quintessence model with $c_s^2=1$.

\begin{figure}
\includegraphics[width=9cm,keepaspectratio]{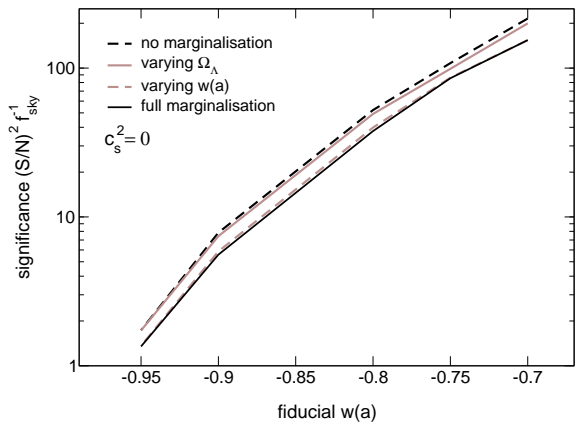}
\caption{The significance of a detection of a $c_s^2=0$ dark energy model compared to a 
$c_s^2=1$ quintessence model as a function of the equation of state, $w(a)$.  
The $S/N$ for a 4-hour integration per pointing is shown
with all parameters fixed to fiducial values (black dashed), 
varying $\Omega_\LLambda$ only (grey solid), varying $w(a)$ 
only (grey dashed), and full marginalisation (black solid).
}\label{signf_w}
\end{figure}

\section{Results} \label{sec:Results}

\subsection{Signal to noise of dark energy sound speed detection}

The basic detection level of a $c_s^2=0$ model relative to a $c_s^2=1$ model is shown 
in Fig.~\ref{signf_w} as a function of the equation of state, $w_a$.
It is clear that the significance of a sound speed detection is larger for models with 
a larger constant equation of state.  This results from the fact that a component with a larger
constant equation of state is more like matter, hence variations in its sound speed produce larger
observable effects. For models with $w_0 \gsim -0.9$ the squared signal to noise exceeds 10 suggesting
that these models can produce a detectable effect. 

Fig.~\ref{signf_w} also illustrates the effect of the covariance between the sound speed
and other dark energy parameters. We note that the full covariance degrades the sensitivity 
to the sound speed but not significantly; in the absence of covariance with all other dark energy 
parameters the sound speed constraints
improve by at most twenty percent. The lack of severe degradation in the sensitivity, that one
would expect to arise from degeneracies between the dark energy parameters, is the result of 
information gained from measuring the auto-correlation and cross-correlation spectra in several
redshift bins. We note that
the sound speed covariance with the equation of state parameters, $w_0$ and $w_a,$
degrades the sensitivity more than the covariance with the density $\Omega_\LLambda,$ a result
that was anticipated from studying the power spectrum derivatives in the previous section. 

\begin{figure}
\includegraphics[width=9cm,keepaspectratio]{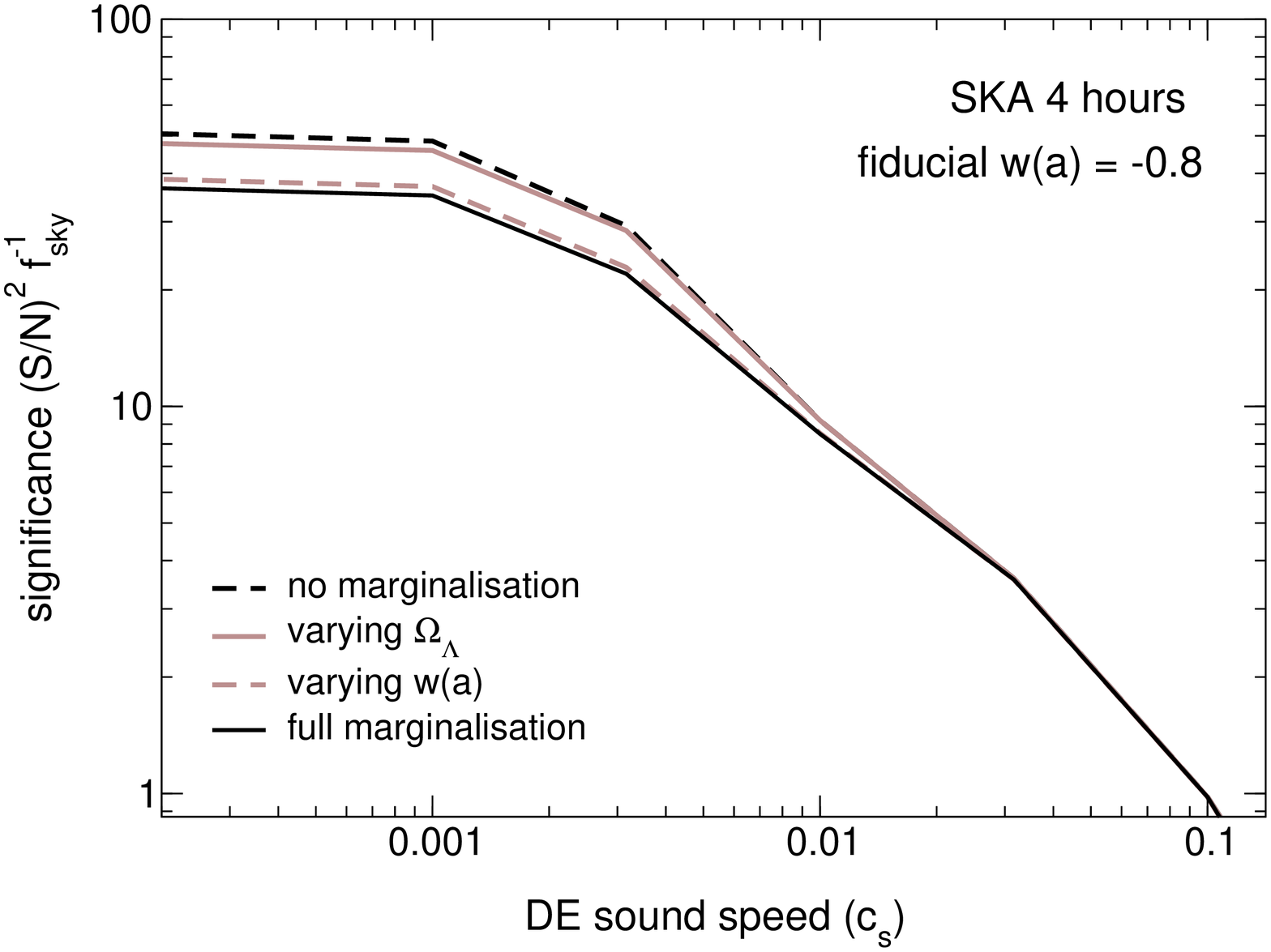}
\caption{The significance of separation between dark energy models with $c_s^2\neq1$ and quintessence 
($c_s^2=1$) for a fiducial $w=-0.8$ model of dark energy. The $S/N$ for a 4-hour integration per 
pointing is shown
with all parameters fixed to fiducial values (black dashed), 
varying $\Omega_\LLambda$ only (grey solid), varying $w(a)$ 
only (grey dashed), and full marginalisation (black solid).} \label{signf_c}
\end{figure}

The significance of detection of dark energy models with $c_s^2 < 1$ relative to a quintessence
model with $c_s^2 = 1$ is shown in Fig.~\ref{signf_c} for a fiducial equation of state, $w(a)=-0.8$.
It is clear that the sound speed detection is most significant for dark energy models with 
$c_s \rightarrow 0$ as the dark energy clustering is most pronounced in these models, and becomes 
marginal for models with $c_s \gsim 0.1$. We also note that the effect of the covariance with other 
dark energy parameters on the sound speed measurement is most evident for models that have a high 
significance of sound speed detection, and hardly noticeable for models in which the sound speed is 
poorly measured.

\begin{figure}
\includegraphics[width=9cm,keepaspectratio]{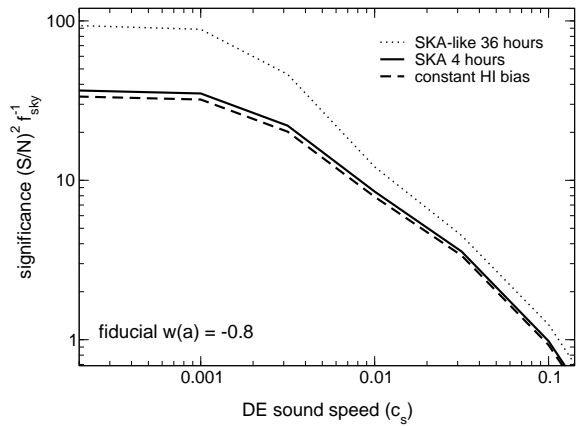}
\caption{The significance of separation between dark energy models with $c_s^2\neq1$ and quintessence 
($c_s^2=1$) for a fiducial $w=-0.8$ model of dark energy.
The marginalised $S/N$ for an SKA 4-hour-per-pointing survey (solid line) and a 
deeper survey of 36 hours per pointing integration time (dotted line). Both
curves above are for an evolving HI bias model. The dashed line corresponds to a constant 
HI bias model.}\label{signf_c_biasdndz}
\end{figure}

We next consider the impact of the uncertainty in our HI bias model on the sound speed constraints.
In Fig.~\ref{signf_c_biasdndz} 
we compare the detection significance as a function of the sound speed for 
our two HI bias models. The constant bias model, which is unlikely to be realistic, 
predicts lower source counts at high redshift. Nevertheless this model 
has a sensitivity to the sound speed that is only about ten percent worse than the 
sensitivity of the evolving bias model. This indicates that the uncertainty in the HI 
bias does not change our forecasts significantly. 

Finally, we have explored how the survey depth affects the significance of sound speed detection.
In Fig.~\ref{signf_c_biasdndz} 
we consider the significance of detection for a 4-hour-per-pointing survey
compared to a 36-hour-per-pointing survey. We have assumed that both surveys cover the same
fraction of the sky so that the 36-hour-per-pointing survey takes nine times longer to 
complete. The longer survey is unlikely to be practical in terms of total integration time but provides
a useful guide as to how significant a much deeper survey will be for constraining the sound
speed. From Fig.~\ref{signf_c_biasdndz} it is clear that the deeper survey is able to discriminate more 
easily between sound speed models. The signal to noise increases 
by a factor of two to three for low sound speed models but is 
approximately the same for models with $c_s \rightarrow 1$. 

\begin{figure} 
\includegraphics[width=9cm,keepaspectratio]{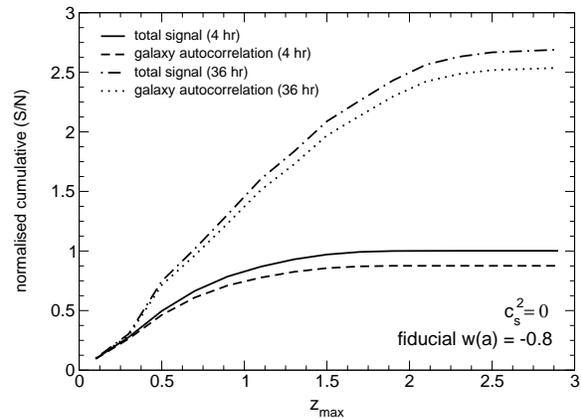}
\caption{The cumulative significance $S/N$ as a function of redshift contributing to
the sum in Eq.~(\ref{eqn:Fisher}) for a $c_s^2=0$ and $w=-0.8$ dark energy model. 
The curves are normalized to the maximum value of the total signal from the 4-hour-per-pointing survey. 
The dashed line represents the signal from the galaxy auto-correlation alone while the solid line represents the total signal from the auto-correlation and cross-correlation spectra. Both curves are for the 4-hour-per-pointing survey. The corresponding 
curves for the deeper 36-hour-per-pointing survey are shown as a dotted line (auto-correlation alone) and as a dot-dashed line (both auto- and cross-correlation).}
\label{fig:cumulative} 
\end{figure}

It is interesting to ask whether this improvement arises from the measurement of clustering
at higher redshifts, $z \gsim 2,$ or from the increased number counts at intermediate redshift.
In Fig.~\ref{fig:cumulative} we plot the cumulative contribution of different redshift bins to the 
discriminating signal for a $c_s^2=0$ and $w=-0.8$ model. 
We note that for the fiducial 4-hour-per-pointing survey nearly all the signal accumulates 
by $z=1.5,$ and only ten percent of the signal comes from the 
cross-correlation spectra.  For the deeper survey, approximately eighty percent of the information comes from $ z \lsim 1.5$. This suggests that the improved signal in Fig.~\ref{signf_c_biasdndz} arises mostly from the increase in the number counts at intermediate redshifts (from $z=0.5$ to $z=1.5$), which results in a lower variance in Eq.~\ref{eqn:Fisher}, rather than the clustering signal at higher redshift.

\section{Conclusions} \label{sec:Conclusions}

We have studied the potential of large HI surveys to constrain constant sound speed
models of dark energy. We investigated the covariance between the dark energy cosmological 
parameters, finding that uncertainties in the density of dark energy and in its equation of 
state will not dramatically degrade our ability to detect the sound speed. This arises because of the 
ability of these surveys to detect large numbers of HI galaxies in several redshift bins. The 
slight reduction in the signal to noise comes mostly from variations in the equation of state 
parameters.

We have also investigated the impact of using an ultra deep SKA-like HI redshift survey and
assessed the effect of changing our HI bias model.
We discovered that a deep survey of HI galaxies, with 36 hours of integration time per pointing
improves constraints on the sound speed as compared with a smaller 4-hour-per-pointing survey 
due to the increased number counts at $z \lsim 1.5$. A maximum redshift depth of $z_{\rm max}\approx 1.5$ 
provides most of the discriminating signal for both surveys. 
In addition, our results have
not shown a strong dependence on the uncertainty in the HI bias model. These results could
guide future planning for these types of survey experiments.

Regarding the detection of the sound speed, we found that we can only detect models of dark
energy with small values of the constant sound speed, $c_s \lsim 0.01$.
As $c_s^2\rightarrow0$, a model with $w=-0.9$ can be detected at the $3$-$\sigma$ level. 
For larger values of $w_0$ sound speeds closer to zero can be detected with greater confidence. The study of the clustering properties of dark energy through its sound speed thus
promises to be an interesting approach to confront the predictions of theoretical dark energy models 
and uncover the nature of this mysterious component.

\section*{Acknowledgements}

ATR acknowledges the SKA project office in South Africa for financial support during his PhD.
KM and CC acknow\-ledge financial support from the National Research Foundation (South Africa).

\label{lastpage}


\end{document}